\documentclass[10pt,aps,prl,reprint,showpacs,superscriptaddress]{revtex4-1}

\usepackage[utf8]{inputenc}	
\usepackage[T1]{fontenc}
\usepackage{color}
\usepackage{amsmath}
\usepackage{amsfonts}
\usepackage{latexsym}
\usepackage{amssymb}
\usepackage{bm}

\usepackage{graphics}
\usepackage{epsfig}
\usepackage{dcolumn}

\usepackage{hyperref}

\newcommand{\g}[1]{{\bf #1}}
\newcommand{\ug}[0]{UGe$_{2}$}

\usepackage{tikz}
\newcommand*\circled[1]{\tikz[baseline=(char.base)]{
            \node[shape=circle,draw,inner sep=2pt] (char) {#1};}}

\begin{document}

\title{Criticalities in the itinerant ferromagnet UGe$_{2}$}
 
\author{Marcin M. Wysoki\'nski}
\email{marcin.wysokinski@uj.edu.pl}

\affiliation{Marian Smoluchowski Institute of Physics, Jagiellonian University, 
\L{}ojasiewicza 11, PL-30-348 Krak\'ow, Poland}

\author{Marcin Abram}
\email{marcin.abram@uj.edu.pl}

\affiliation{Marian Smoluchowski Institute of Physics, Jagiellonian University, 
\L{}ojasiewicza 11, PL-30-348 Krak\'ow, Poland}

\author{J\'ozef Spa\l ek}
\email{ufspalek@if.uj.edu.pl}

\affiliation{Marian Smoluchowski Institute of Physics, Jagiellonian University, 
\L{}ojasiewicza 11, PL-30-348 Krak\'ow, Poland}
\affiliation{Academic Centre for Materials and Nanotechnology, AGH University of Science and Technology, Aleja Mickiewicza 30, PL-30-059 Kraków, Poland}

\begin{abstract}
We provide a microscopic description of the magnetic properties of \ug\ and in particular,
of its both classical and quantum critical behavior.
Namely, we account for all the critical points:
the critical ending point (CEP) at the metamagnetic phase 
transition, the tricritical point, and the quantum critical
end point at the ferromagnetic to paramagnetic phase transition.
Their position agrees quantitatively with experiment.
Additionally, we predict that the metamagnetic CEP can be traced down to 
zero temperature and becomes quantum critical point by a small decrease of both the 
total electron concentration and the external pressure.
The system properties are then determined by the quantum critical 
fluctuations appearing near the instability point of the Fermi surface topology.

\end{abstract}


\maketitle

{\it Introduction.}
Attempts to determine the quantum critical behavior and  the corresponding critical points (QCPs) 
have attracted much attention due to the unique phenomena with singular physical
properties associated with them as temperature $T \rightarrow 0$ 
and other parameters (pressure $p$, applied field $H$, or electron concentration $n$)
are varied \cite{Lohneysen2007, Carr2011,Slebarski2005}.
Additionally, in the canonical case---the heavy fermion systems---unconventional superconductivity 
often appears near those QCPs making the quantum critical fluctuations 
the primary pairing inducing factor.
Also, the classical critical points (CCPs) and their evolution towards QCP
provide the testing ground for study of detailed quantitative behavior of different systems
\cite{Pfleiderer2009,Spalek1987, *SpalekPhysStatSol2006}.

\ug, in this respect, is one of the unique materials that exhibit
all the above features.
Therefore, the explanation of the magnetic phase diagram and intimately 
connected critical points within a single theoretical framework 
would provide a complete understanding of this remarkable quantum material 
\cite{Saxena2000,Pfleiderer2002,Pfleiderer2009,Taufour2010,Kotegawa2011}.
The phase diagram on the pressure--temperature ($p$--$T$) plane comprises
two ferromagnetic phases, of weaker (FM1) and stronger (FM2) magnetization, 
paramagnetic phase (PM), as well as the spin-triplet 
superconducting phase (SC) \cite{Saxena2000,Huxley2001,Pfleiderer2009}.
SC disappears at the same pressure as FM \cite{Saxena2000} 
and the maximum of the superconducting critical temperature $T_s$ coincides with the critical pressure for 
the FM2-FM1 phase transition \cite{Pfleiderer2002}.
Thus, it is strongly suggestive that FM and SC are strongly intertwined 
as described by some theoretical approaches \cite{Kirkpatrick2001,Machida2001,Abrikosov2001,Sa2002,Sandeman2003}.

The $p$-$T$-$H$ phase diagram for \ug\ comprises the
characteristic {\it wing shape} 
\cite{Taufour2010,Kotegawa2011}. Such structure was
theoretically obtained by Belitz et al. \cite{Kirkpatrick2005}
within mean-field approach for a single-band itinerant ferromagnet.
However, this approach cannot account for the two different ferromagnetic phases
appearing in \ug, as well as for the critical ending point (CEP), separating 
the region with a discontinuous drop in magnetization from a crossover regime \cite{Hardy2009, Taufour2010}. 
 
In this work we provide a quantitative microscopic description 
of all magnetic critical properties of \ug\ within 
the framework of the Anderson lattice model (ALM) treated by a modified Gutzwiller 
approach \cite{Rapid}, called {\it the statistically consistent 
Gutzwiller approximation} (SGA) (for a description of the 
method and a detailed comparison to the slave-boson
approach see Ref. \cite{sga}; for its applications, see
Refs. \cite{Jedrak2011, *Kaczmarczyk2011,*Howczak2013,
*Kadzielawa2013,*Abram2013,*Zegrodnik2013,*Wysokinski2014}).
Validity of this model in the context of \ug\ \cite{Rapid} is 
based on earlier results: first, on band structure calculations \cite{Shick2001, Samsel2011}
and second, on experimental observations \cite{Saxena2000,Tran2004, Pfleiderer2009}. 
The first feature is a quasi-two-dimensional topology of the 
Fermi surface (FS) \cite{Shick2001, Samsel2011} which justifies calculations for a two-dimensional square lattice.
On the other hand, despite the circumstance that the distance between uranium
atoms is above the Hill limit \cite{Pfleiderer2009}, the experimental value of the paramagnetic moment per U atom
is different from that for either $f^3$ or $f^2$ configurations \cite{Saxena2000,Kernavanois2001}. This speaks for 
the presence of a sizable hybridization between the initially localized $f$ electrons and those from
the conduction band. For strong enough hybridization, $f$ electrons contribute essentially 
to the heavy itinerant quasiparticle states and play a dominant role in the magnetic properties 
\cite{Saxena2000, Huxley2001, Kernavanois2001}.  

We provide a coherent explanation of FM and 
PM phase appearances as driven by a competition between the 
hybridization from one side and the 
$f$--$f$ Coulomb local repulsive interaction from the other \cite{Rapid}. 
Specifically, we obtain two  different FM
phases \cite{Doradzinski1997,Doradzinski1998,Sandeman2003,Howczak2012,Kubo2013,Rapid}
by varying the predetermined position of the chemical potential 
with respect to the peaks in the quasiparticle density of 
states (DOS) including the spin-split subbands.
Although, Gutzwiller ansatz in certain regimes favors 
antiferromagnetism over FM
\cite{Kotliar1986,Dorin1992,Doradzinski1997,Doradzinski1998,Howczak2012},
we restrict our discussion to the latter phase, because 
in the considered range of electron concentration, $n\simeq 1.6$, 
FM phase turned out to have the lowest energy 
\cite{Doradzinski1997,Doradzinski1998}.

\begin{figure}[t]
  \centering
  \includegraphics[width=0.451\textwidth]{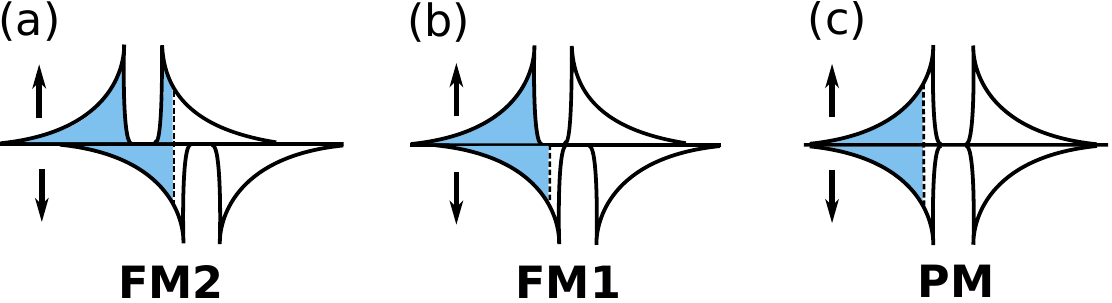}
   \caption{(Color online) Schematic characterization of phases by their 
   spin-resolved density of states. The arrows label the spin 
   subbands and the dotted line marks the position of the chemical potential. }\label{fig0}\vspace{-0.35cm}
  \end{figure}

In Fig.~\ref{fig0} we 
draw schematically the respective DOS for considered phases.  
It can be seen clearly that the shape of the FS (limiting the filled parts) 
will be vastly different in each of the phases. 
Within our approach, most of the 
properties of \ug\ at $T=0$ can be explained \cite{Rapid} 
in agreement with related experiments of
magnetization \cite{Pfleiderer2002}, 
neutron scattering \cite{Kernavanois2001,Huxley2001}, 
and the de Haas--van Alphen oscillations \cite{Terashima2001,Settai2002}.
The character of the FM1 phase, which we obtain as a
half-metallic type [cf.\ Fig.\ \ref{fig0}(b)], is also supported by 
the band-structure calculations \cite{Samsel2011}.

 In the present work we extend our previous approach \cite{Rapid} to nonzero temperature
and on this basis we determine the character of all phase
transitions on the $p$--$T$--$H$ diagram of \ug, as well as discuss the nature
of all the classical and quantum critical points. 
We also show that by a small decrease 
of electron concentration (by $\sim{}7\%$), the system can 
reach another quantum criticality via a metamagnetic transition upon changing the pressure.
We also predict the corresponding change in FS topology distinguishing the two phases 
of significantly different magnetic susceptibility.

{\it Model.}
We start from ALM with the Zeeman term included
($h \equiv \frac{1}{2} g \mu_0\mu_B H$) in the Hamiltonian
\begin{equation}
 \begin{split}
 \mathcal{\hat H}-\mu\hat N=&{\sum_{\g i,\g j,\sigma}}'  t_{\g i \g j}\hat c_{\g i,\sigma}^\dagger\hat c_{\g j,\sigma}
-\sum_{\g i,\sigma}(\mu+\sigma h)\hat n^c_{\g i,\sigma}\\
&+ \sum_{\g i,\sigma}(\epsilon_f-\mu-\sigma h)\hat n^f_{\g i,\sigma}+ 
U\sum_{\g i} \hat n^f_{\g i,\uparrow} \hat n^f_{\g i,\downarrow}\\
&+ V\sum_{\g i,\sigma}(\hat f_{\g i,\sigma}^\dagger
\hat c_{\g i,\sigma}+\hat c_{\g i,\sigma}^\dagger\hat f_{\g i,\sigma}),\label{Ho}
\end{split}
\end{equation}
which comprises dispersive conduction ($c$) 
band electrons and $f$ electrons coming from atomic $f$-shell located at $\epsilon_f<0$. 
In the model we include specifically the nearest ($t<0$) and
the second nearest ($t'=0.25|t|$) neighbor hopping amplitudes
between $c$ electrons, $f$ level at $\epsilon_f=-3|t|$,
sizable $f$--$f$ Coulomb repulsion $U=5|t|$, and the $c$--$f$ hybridization $V$ of the on-site form.

To obtain an effective single particle 
picture from the many-body Hamiltonian (\ref{Ho}) 
we use the extended Gutzwiller approximation (GA) 
 called the SGA 
(for details see \cite{sga}). The method was successfully applied to 
a number of problems \cite{Jedrak2011,*Kaczmarczyk2011,*Howczak2013,
*Kadzielawa2013,*Abram2013,*Zegrodnik2013,*Wysokinski2014}. 
Formally, we add to the effective Hamiltonian obtained in GA \cite{Rice1985, Fazekas1987}, 
$\mathcal{\hat H}_{GA}$, additional 
constraints on the number of $f$ electrons and their magnetization 
by means of the Lagrange multipliers. 
It leads to the new effective Hamiltonian
$\mathcal{\hat H}_{SGA}$ of the form,
\begin{multline}
 \mathcal{\hat H}_{SGA} \equiv  \\
 \mathcal{\hat H}_{GA} - \lambda^f_n \Big( \sum_{\g{k},\sigma}\hat n^f_{\g{k},\sigma}- \Lambda n_f \Big)
 -\lambda^f_m \Big( \sum_{\g{k},\sigma}\sigma\hat n^f_{\g{k},\sigma} -  \Lambda m_f \Big) \vspace{3pt}\\
\equiv \sum_{\g{k},\sigma}  \hat\Psi^\dagger_{\g{k}\sigma}
\begin{pmatrix}
 \epsilon_{\g{k}}^{c}-\sigma h-\mu&\sqrt{q_\sigma}V \vspace{3pt}\\
 \sqrt{q_\sigma}V& \epsilon_{f}-\sigma (h+\lambda_{m}^f)-\lambda_{n}^f-\mu   \\ 
\end{pmatrix}
\hat\Psi_{\g{k}\sigma} \\
+\Lambda( Ud^2 +\lambda_{n}^fn_f +\lambda_{m}^fm_f),
\label{HSGA}
\end{multline}
where $\hat\Psi_{\g{k}\sigma}^\dagger\equiv(\hat c_{\g{k},\sigma}^\dagger, \hat f_{\g{k},\sigma}^\dagger)$.
Furthermore, $q_\sigma$ is the hybridization narrowing factor 
in the standard form \cite{Rapid,Wysokinski2014},
and $\Lambda$ is a number of lattice sites.

At nonzero temperature,
one needs to minimize the generalized Landau grand potential functional
\begin{equation}
\begin{split}
\frac{ \mathcal{F}}{\Lambda}  = & -\frac{1}{\Lambda\beta} \sum_{{\bf k} \sigma b} \ln[1+e^{-
\beta E_{{\bf k}\sigma}^b}]\\
&+ (\lambda_{n}^fn_f+\lambda_{m}^fm_f+Ud^2),
\label{5}
\end{split}
\end{equation}
where $E_{\g k\sigma}^b$ are four eigenvalues of the effective Hamiltonian (\ref{HSGA})
labeled with the spin ($\sigma$) and band ($b$) indices.
$\lambda_n^f$ and $\lambda_m^f$ are the Lagrange multipliers assuring the correct
statistical consistency of equations for $n_f$ and $m_f$ and play the role of 
correlation-induced effective fields \cite{Wysokinski2014}.
Minimization of $\mathcal{F}$ is carried out with respect
to the set of all parameters $\vec \lambda\equiv\{d,n_f,m_f,\lambda_n^f,\lambda_m^f\}$.
Additionally, as the number of particles in the system is conserved we have to 
determine the chemical potential and adjust it to each of the phases according 
to the condition $n=1/\Lambda\sum_{{\bf k}b\sigma}  f ( E_{{\bf k} \sigma}^{b} )$ ,
with $f(E)$ being the Fermi-Dirac function. In effect, the model is described by set of
six algebraic equations which are solved
with the help of the GSL library, with typical accuracy $10^{-11}$. 

The Landau grand-potential functional for the equilibrium values of the parameters, $\mathcal F_0$,
has the meaning of the physical grand-potential $\Omega$ 
which is the proper quantity for studying the system at any temperature,
$\mathcal F_0\equiv\Omega\equiv U-TS-\mu N$.
Therefore, the free energy of the system 
is defined by $F= \mathcal F_0 + \mu N$ and the ground-state energy is $E_G\equiv F(T=0)$.

  \begin{figure}
  \centering
  \includegraphics[width=0.5\textwidth]{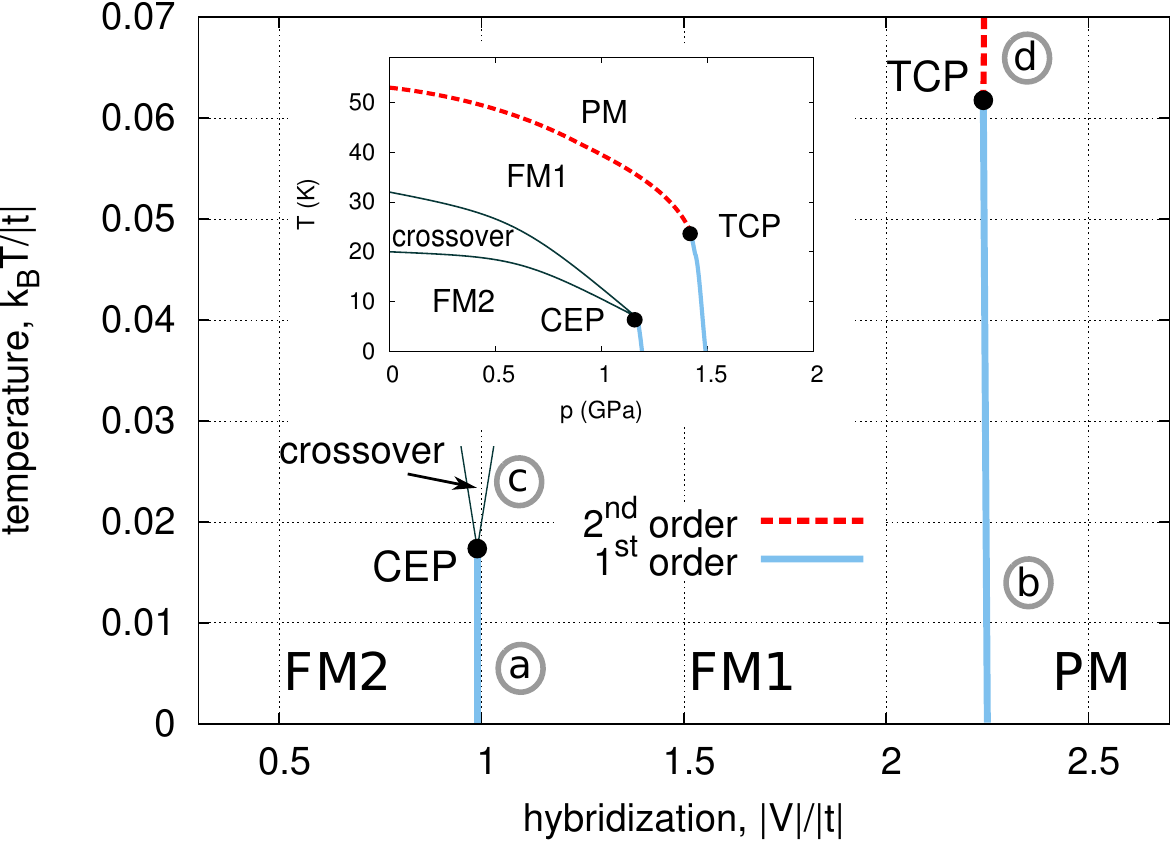}\\
  \includegraphics[width=0.25\textwidth]{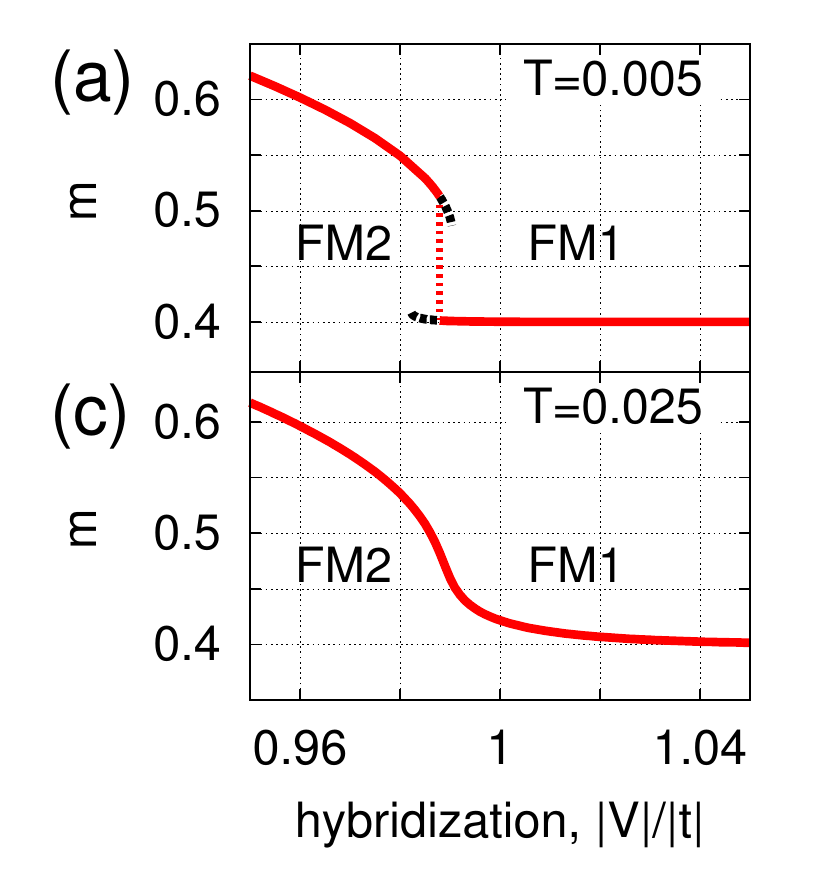} \hspace{-5mm}
  \includegraphics[width=0.25\textwidth]{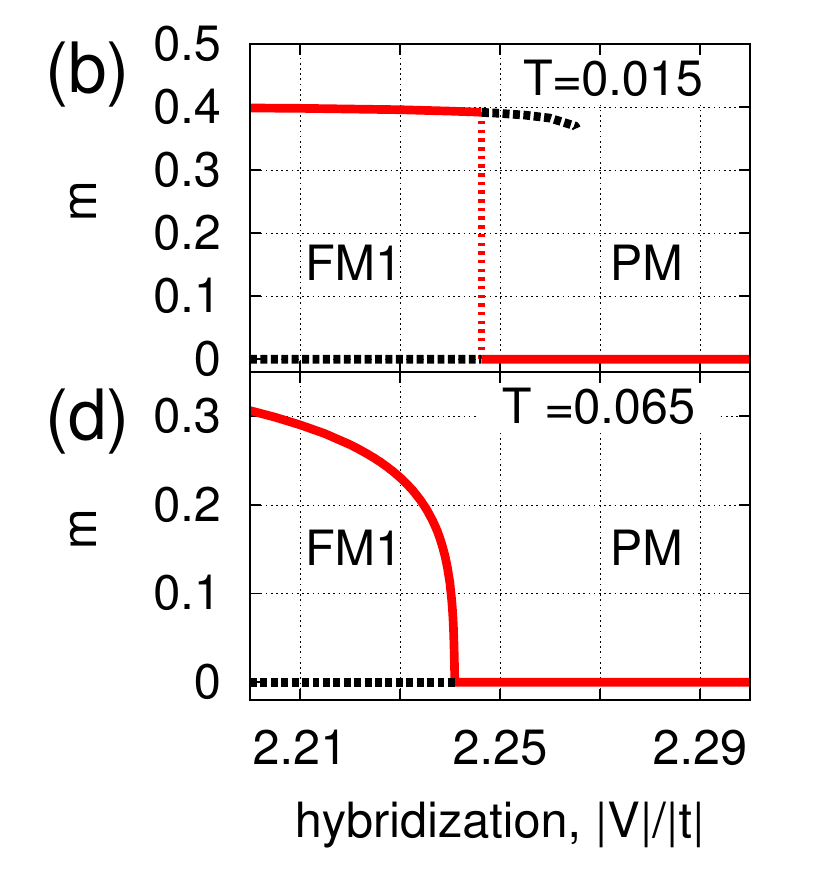}\hspace{-5mm}
   \caption{(Color online) Top: Phase diagram on 
   hybridization strength--reduced temperature plane
   encompassing both FM and PM phases for total band filling $n=1.6$. 
   The correct character of phase transitions and 
   positions of critical points in \ug\ 
   \cite{Pfleiderer2002,Hardy2009,Taufour2010,Kotegawa2011} 
   is reproduced. For comparison, we
   present in the inset the experimental 
   $p$--$T$ phase diagram of \ug\ (cf. \cite{Pfleiderer2002,Taufour2010}).
   In (a)--(d) we draw the magnetization change with the increasing hybridization
   strength when the system undergoes phase transition at points 
   indicated with respective encircled letters \protect\circled{a}--\protect\circled{d}. Solid red lines denote energetically 
   favorable solution, whereas dashed black lines denote the unstable solutions.}\label{fig1}\vspace{-0.35cm}
  \end{figure}

 {\it Results.} We assume that the main 
effect of the applied pressure is emulated by an increase 
of the hybridization amplitude $|V|$, even though other parameters (e.g., $\epsilon_f$)
may also change. However, as our previous results indicate, hybridization change is the principal
factor of the pressure dependencies observed in \ug\ \cite{Rapid}.

In Fig.~\ref{fig1} we plot the phase diagram on the $|V|$--$T$ plane. In the low-$T$
regime we are able to reproduce the correct evolution of both
metamagnetic (left) and ferromagnetic to paramagnetic (right) phase transitions 
observed in experiment (cf.~inset), together with the respective critical behavior 
\cite{Pfleiderer2002,Hardy2009,Taufour2010,Kotegawa2011}. 
The position of the classical critical points (CCPs) is very sensitive to
the selected total band filling, $n=n_f+n_c$. 
Our fitting constraint is the ratio of the corresponding critical temperatures,
$T_{CEP}/T_{TCP}\approx7K/24K$ \cite{Taufour2010}.
Consequently, for the band filling $n=1.6$, 
selected in our previous analysis 
at $T=0$ \cite{Rapid}, we obtain agreement of our
calculated ratio under the proviso that experimental values of the critical 
temperatures are determined with accuracy $\pm0.25K$.

Our model does not account for correct curvatures of phase transitions above 
CCPs (cf. Fig.~\ref{fig1}). 
 This discrepancy can be attributed to the fact that also other microscopic parameters 
can alter when applying pressure, e.g. $\epsilon_f$, and to 
additional entropic factors important in the case of $T>0$ Gutzwiller projection 
\cite{Wang2010, Fabrizio2013}.

In our calculations we have used 
reduced temperature $k_BT/|t|$. We rescale it to the 
physical units by relating it to the experimentally measured values at CCPs \cite{Pfleiderer2002,Hardy2009,Taufour2010,Kotegawa2011}.
Accordingly, we also rescale the reduced field
$\frac{1}{2} g \mu_B\mu_0 H/|t|$ to Tesla units. 

At the metamagnetic (FM2-FM1) phase transition we obtain CEP 
separating the discontinuous-transition line from the crossover regime.
At low T both solutions 
with the weaker and the stronger magnetization coexist in the limited 
range of the hybridization strength [cf.\ Fig.~\ref{fig1}a]. As the system 
approaches the transition from the FM1 side, FS changes 
drastically only in one spin-subband, in which 
the chemical potential crosses the hybridization gap, resulting also 
in a discontinuous jump of the total moment $m=m_f+m_c$. 
With the increasing temperature, the edges of the gap are 
gradually smeared out. This leads to a deviation from the pure half-metallic
type of the FM1 phase. The magnetization is {\it bending} towards
the trend observed in the FM2 phase, and eventually at CEP it is changing to a
crossover line [cf.\ Fig.\ \ref{fig1}(c)]. 

  \begin{figure}
  \centering
  \includegraphics[width=0.5\textwidth]{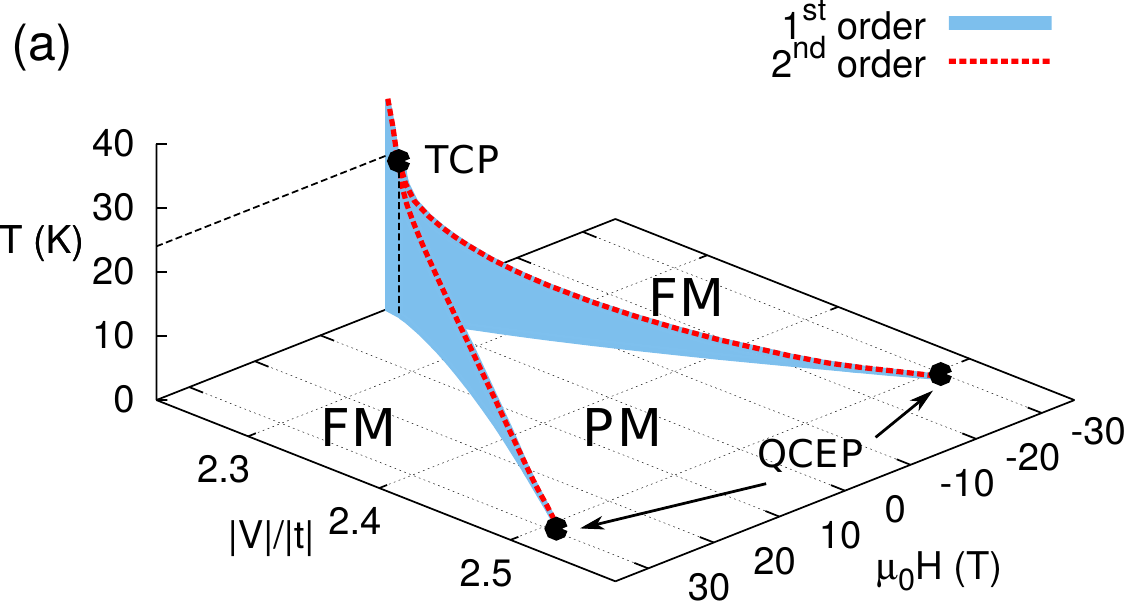}\vspace{0.5cm}
  \includegraphics[width=0.49\textwidth]{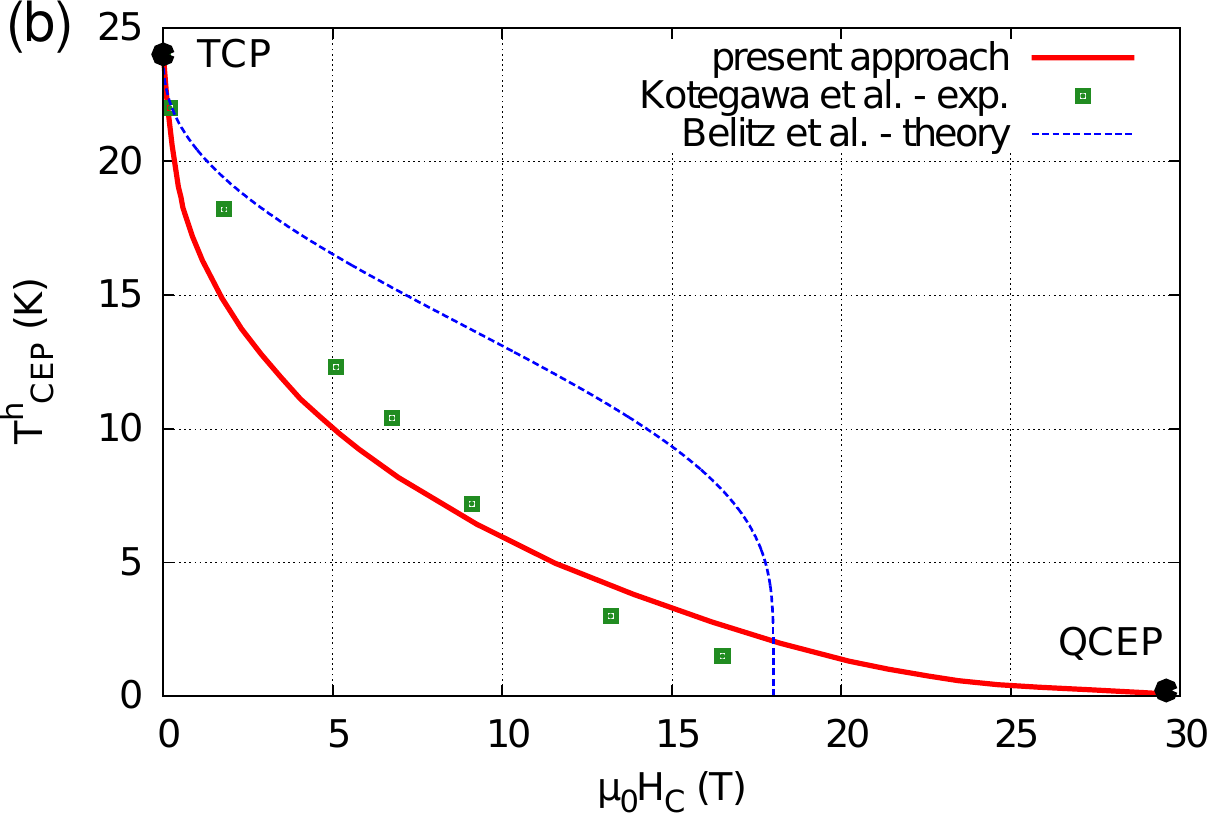}
   \caption{(Color online) (a) Wing structure of the phase 
   transition planes derived from our model. (b) Comparison
   of the calculated dependence of the temperature vs applied magnetic field at
   the critical end point (CEP) with the experimental points adopted from \cite{Kotegawa2011}.
  For comparison, we include also the prediction by Belitz et al.~\cite{Kirkpatrick2005},
  with the fitting parameters selected on the basis \cite{Kotegawa2011}:
  $H_{QCEP}=18$~T and $T_{TCP}=24$~K.}\label{fig2}\vspace{-0.35cm}
  \end{figure}

In the case of FM to PM transition 
the situation is different [cf. Figs. \ref{fig1}(b), and \ref{fig1}(d)].
At low temperature, the magnetization of this 
{\it half-metallic} FM1 phase discontinuously drops to zero (cf.\ Fig.\ \ref{fig1}b).
However, with the increasing temperature, the ferromagnetic solution departs from
a sharp half-metallic type and slowly {\it bends} over towards the paramagnetic
solution, eventually reaching the critical point by changing the transition character
to that of second order [cf.\ Fig.\ \ref{fig1}(d)]. 
The just described critical point is of tricritical character (TCP). This is because its evolution 
can be followed by applying the magnetic field down to $T=0$,
where it turns into the quantum critical ending point (QCEP) (cf.\ Fig.\ \ref{fig2}a).
In this manner, we have achieved a full characteristic at the {\it wing-shape} 
$p$--$T$--$H$ phase diagram \cite{Taufour2010,Kotegawa2011}. 
As the detailed form of the hybridization change with 
applied pressure is unknown, and in principle non-linear, we compare
our predicted shape of wings by tracing the evolution of CEP on the 
temperature---magnetic field $T^h_{CEP}$ -- $\mu_0H_C$ plane 
[cf.\ Fig.\ \ref{fig2}(b)] and comparing it to the experimental
data \cite{Kotegawa2011}. We obtain a satisfactory 
quantitative agreement with the experimental points, as well as
recover its proper curvature.
For comparison, the results from the mean-field approach 
to the single-band case by Belitz et al. \cite{Kirkpatrick2005} are also drawn, 
as is universal explanation of tricritical behavior of itinerant ferromagnets.
Nevertheless, as suggested by the authors in Ref. \cite{Kotegawa2011}, the crucial 
element determining for \ug\ the correct shape of the wings is the change of FS, 
present in our two-band ALM model. 
We also predict that the curve of the $T^h_{CEP}$ vs $\mu_0H_C$ dependence
has a longer tail than that estimated in Ref.~\cite{Kotegawa2011}, i.e., that 
QCEP should be located at fields around $30$~T. 
Our estimate thus calls for a more precise determination of the QCEP position.

  \begin{figure}
   \centering
  \includegraphics[width=0.5\textwidth]{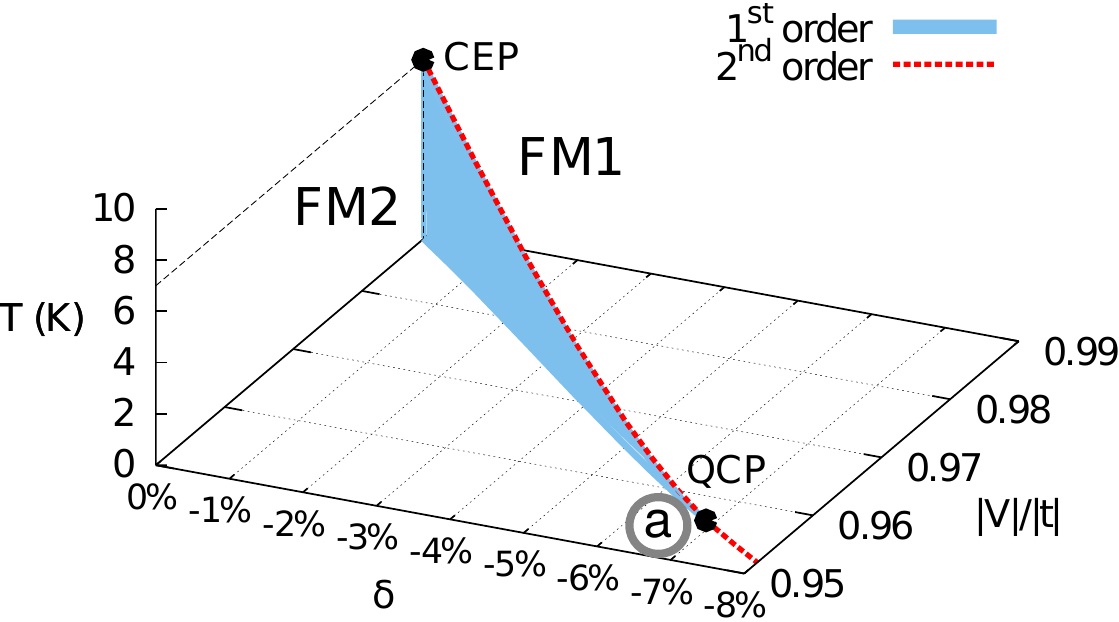}\vspace{6pt}
    \includegraphics[width=0.215\textwidth]{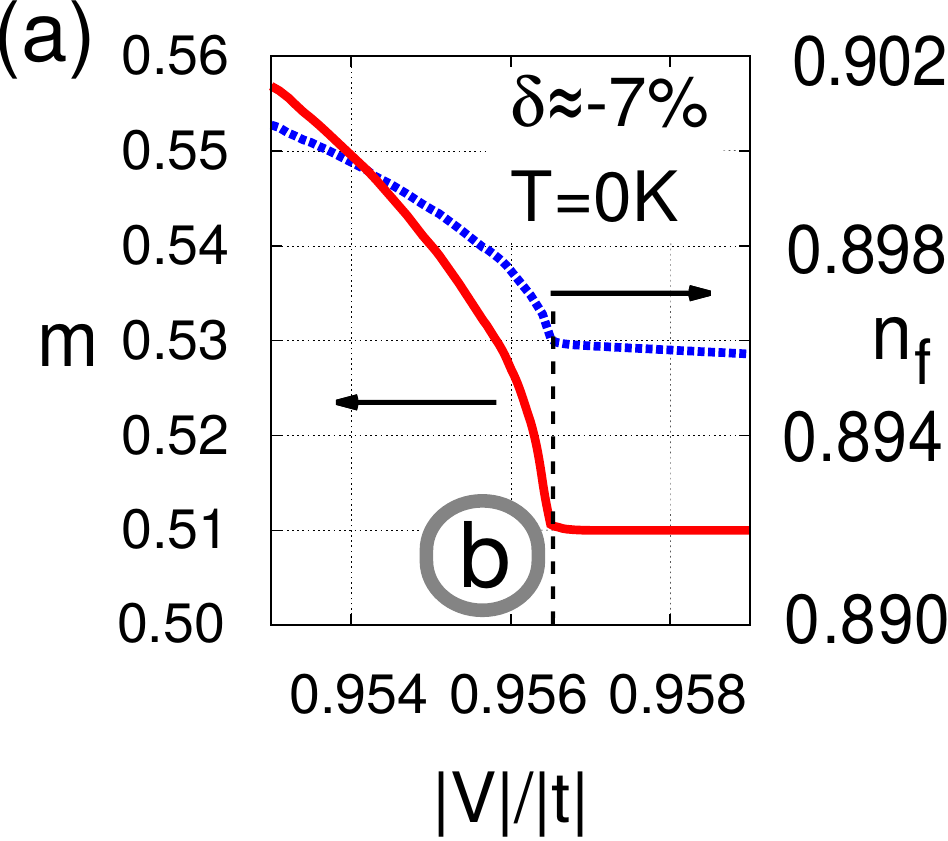}\hspace{5pt}
  \includegraphics[width=0.25\textwidth]{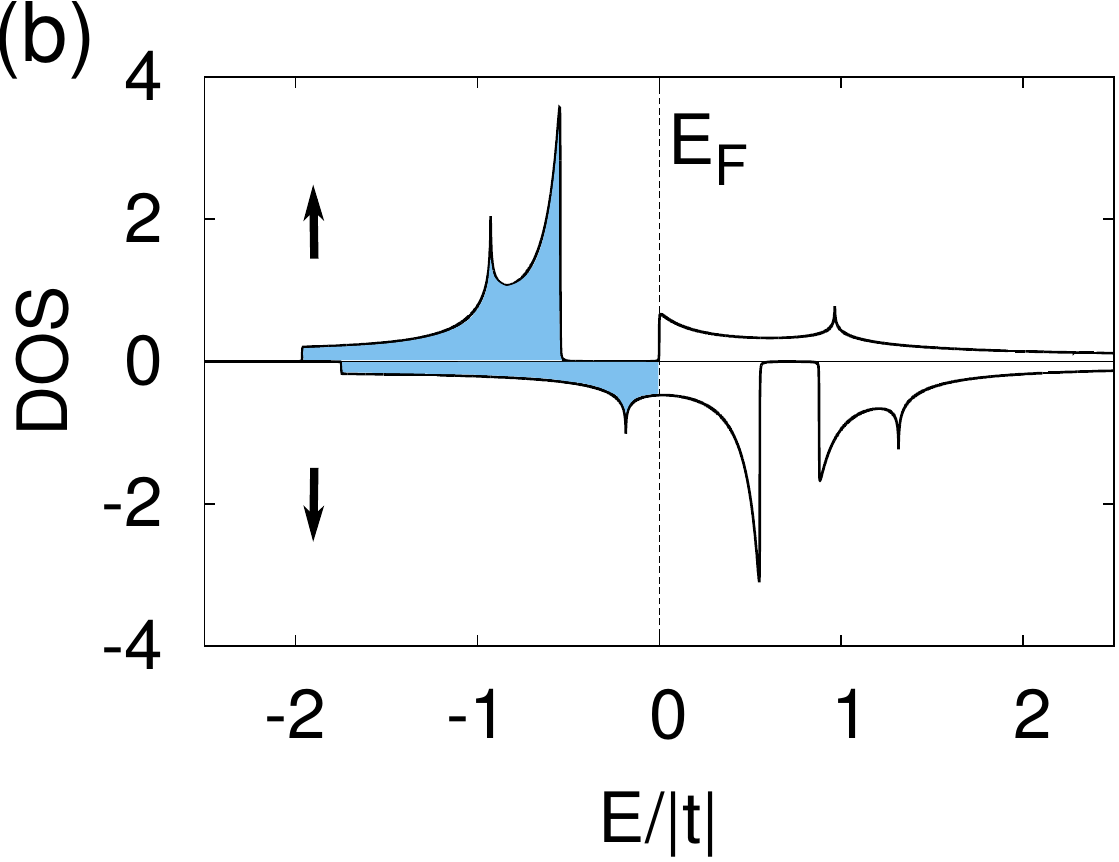}
   \caption{(Color online) Top: Evolution of CEP on the
   $|V|$--$T$--$\delta$ plane down to $T=0$ and QCP (see main text).
   Bottom: (a) Change of magnetization and $f$ electron number as the system 
   undergoes quantum critical transition. (b) Density of states at QCP.
   Note the intermediate character of FS between FM2 and FM1 of the state at QCP.
   Encircled letter \protect\circled{a} at top diagram refers to the position of the curves in panel (a), and
respectively \protect\circled{b} at panel (a) to the position of the DOS in~(b).}\label{fig3}\vspace{-0.35cm}
  \end{figure}
 
In fitting to the data in Fig. \ref{fig2} we have assumed that the 
$g$-factor for $f$ electrons $g_f=2$ (the same as for $c$~electrons).
This assumption is based on the presumption that for itinerant electrons the crystal-field multiplet  structure is washed out.
Parenthetically, taking $g_f$ significantly different provides a worse agreement, but the curvature character remains unchanged.
 
In Fig.\ \ref{fig3} we draw the evolution of CEP 
at the metamagnetic transition with the decrease of both the 
hybridization and the electron concentration. 
The latter quantity is characterized by 
the parameter $\delta=\frac{n_x-n}{n}100\%$, where $n=1.6$
is initial and $n_x$ is the actual concentration.  
On the $V$--$T$--$\delta$ phase diagram 
the CEP can be followed down to zero temperature, where it joins the
second-order transition line [cf.\ Fig.\ \ref{fig3}(a)].
At this second order transition the Fermi level for the majority spin subband is 
exactly at the border of the gap [cf.\ Fig.\ \ref{fig3}(b)]. It means 
that along this line quantum critical fluctuations of FS topology are present. 
In other terms, we have a strong indication that in the vicinity of the SC dome maximum this compound exhibits
a {\it Lifshitz} type of quantum critical behavior. 
This quantum critical transition can be associated also 
with the specific valence change [cf.\ Fig.\ \ref{fig3}(a)]. 
However, here the average $f$ electron number changes continuously in contrast
to the discontinuous drop originating from the $f$--$c$ electron repulsion~\cite{Miyake2014}.
The difference in the origin of {\it Lifshitz} type of 
ferromagnetic QCP with respect to that considered before 
\cite{Millis2001,Fay1980} is that here it results 
from the two-band model and separates different FM phases.

{\it Summary.} We have described the phase diagram of \ug\ at nonzero temperature
and have determined the location of the critical points, 
as well as proposed an additional quantum critical point
for \ug. 
With the help of the Anderson lattice model we are able to reproduce quantitatively
all the principal features of the magnetism in this compound.
We also have determined the location of experimentally observed critical and quantum critical
points, together with a correct order of the phase transitions related to them. 

Although our mean-field approach seems to capture all the features concerning
details of the $p$-$T$-$H$ phase diagram of \ug, we should note that, in principle, fluctuations
of order parameters can bring a quantitative changes to our results. 
However, as the phase transitions are induced by the
drastic changes of the Fermi surface, the effect of the fluctuations should be
minor (except near the predicted QCP---cf. Fig.\ref{fig3}) and may lead to a correction
of the CEP and TCP positions.

It should be noted that we have employed an 
orbitally nondegenerate ALM. Accounting for the degenerate one would imply
inclusion of the residual Hund's rule interaction present in the degenerate ALM model
which could be important in inducing the spin-triplet pairing 
\cite{Zegrodnik2014,*Zegrodnik2013JPCM}.

{\it Acknowledgments.} The work was partly supported by the Foundation for 
Polish Science (FNP) under the Grant TEAM and partly by the  
National Science Centre (NCN) under the MAESTRO, Grant No. DEC-2012/04/A/ST3/00342.
Access to the supercomputer located at Academic Centre for 
Materials and Nanotechnology of the AGH University of Science and Technology in Krak\'ow is also acknowledged.


\begin{thebibliography}{48}%
\makeatletter
\providecommand \@ifxundefined [1]{%
 \@ifx{#1\undefined}
}%
\providecommand \@ifnum [1]{%
 \ifnum #1\expandafter \@firstoftwo
 \else \expandafter \@secondoftwo
 \fi
}%
\providecommand \@ifx [1]{%
 \ifx #1\expandafter \@firstoftwo
 \else \expandafter \@secondoftwo
 \fi
}%
\providecommand \natexlab [1]{#1}%
\providecommand \enquote  [1]{``#1''}%
\providecommand \bibnamefont  [1]{#1}%
\providecommand \bibfnamefont [1]{#1}%
\providecommand \citenamefont [1]{#1}%
\providecommand \href@noop [0]{\@secondoftwo}%
\providecommand \href [0]{\begingroup \@sanitize@url \@href}%
\providecommand \@href[1]{\@@startlink{#1}\@@href}%
\providecommand \@@href[1]{\endgroup#1\@@endlink}%
\providecommand \@sanitize@url [0]{\catcode `\\12\catcode `\$12\catcode
  `\&12\catcode `\#12\catcode `\^12\catcode `\_12\catcode `\%12\relax}%
\providecommand \@@startlink[1]{}%
\providecommand \@@endlink[0]{}%
\providecommand \url  [0]{\begingroup\@sanitize@url \@url }%
\providecommand \@url [1]{\endgroup\@href {#1}{\urlprefix }}%
\providecommand \urlprefix  [0]{URL }%
\providecommand \Eprint [0]{\href }%
\providecommand \doibase [0]{http://dx.doi.org/}%
\providecommand \selectlanguage [0]{\@gobble}%
\providecommand \bibinfo  [0]{\@secondoftwo}%
\providecommand \bibfield  [0]{\@secondoftwo}%
\providecommand \translation [1]{[#1]}%
\providecommand \BibitemOpen [0]{}%
\providecommand \bibitemStop [0]{}%
\providecommand \bibitemNoStop [0]{.\EOS\space}%
\providecommand \EOS [0]{\spacefactor3000\relax}%
\providecommand \BibitemShut  [1]{\csname bibitem#1\endcsname}%
\let\auto@bib@innerbib\@empty
\bibitem [{\citenamefont {L\"ohneysen}\ \emph {et~al.}(2007)\citenamefont
  {L\"ohneysen}, \citenamefont {Rosch}, \citenamefont {Vojta},\ and\
  \citenamefont {W\"olfle}}]{Lohneysen2007}%
  \BibitemOpen
  \bibfield  {author} {\bibinfo {author} {\bibfnamefont {H.~v.}\ \bibnamefont
  {L\"ohneysen}}, \bibinfo {author} {\bibfnamefont {A.}~\bibnamefont {Rosch}},
  \bibinfo {author} {\bibfnamefont {M.}~\bibnamefont {Vojta}}, \ and\ \bibinfo
  {author} {\bibfnamefont {P.}~\bibnamefont {W\"olfle}},\ }\href {\doibase
  10.1103/RevModPhys.79.1015} {\bibfield  {journal} {\bibinfo  {journal} {Rev.
  Mod. Phys.}\ }\textbf {\bibinfo {volume} {79}},\ \bibinfo {pages} {1015}
  (\bibinfo {year} {2007})}\BibitemShut {NoStop}%
\bibitem [{\citenamefont {Si}\ \emph {et~al.}(2011)\citenamefont {Si},
  \citenamefont {Gegenwart},\ and\ \citenamefont {Steglich}}]{Carr2011}%
  \BibitemOpen
  \bibfield  {author} {\bibinfo {author} {\bibfnamefont {Q.}~\bibnamefont
  {Si}}, \bibinfo {author} {\bibfnamefont {P.}~\bibnamefont {Gegenwart}}, \
  and\ \bibinfo {author} {\bibfnamefont {F.}~\bibnamefont {Steglich}},\
  }\enquote {\bibinfo {title} {{\it Understanding Quantum Phase
  Transitions}},}\ \ (\bibinfo  {publisher} {CRC Press, Boca Raton, edited by
  L. D. Carr},\ \bibinfo {year} {2011})\ Chap.\ \bibinfo {chapter} {8, 18},
  pp.\ \bibinfo {pages} {193--216, 445--468}\BibitemShut {NoStop}%
\bibitem [{\citenamefont {\ifmmode~\acute{S}\else \'{S}\fi{}lebarski}\ and\
  \citenamefont {Spa\l{}ek}(2005)}]{Slebarski2005}%
  \BibitemOpen
  \bibfield  {author} {\bibinfo {author} {\bibfnamefont {A.}~\bibnamefont
  {\ifmmode~\acute{S}\else \'{S}\fi{}lebarski}}\ and\ \bibinfo {author}
  {\bibfnamefont {J.}~\bibnamefont {Spa\l{}ek}},\ }\href {\doibase
  10.1103/PhysRevLett.95.046402} {\bibfield  {journal} {\bibinfo  {journal}
  {Phys. Rev. Lett.}\ }\textbf {\bibinfo {volume} {95}},\ \bibinfo {pages}
  {046402} (\bibinfo {year} {2005})}\BibitemShut {NoStop}%
\bibitem [{\citenamefont {Pfleiderer}(2009)}]{Pfleiderer2009}%
  \BibitemOpen
  \bibfield  {author} {\bibinfo {author} {\bibfnamefont {C.}~\bibnamefont
  {Pfleiderer}},\ }\href {\doibase 10.1103/RevModPhys.81.1551} {\bibfield
  {journal} {\bibinfo  {journal} {Rev. Mod. Phys.}\ }\textbf {\bibinfo {volume}
  {81}},\ \bibinfo {pages} {1551} (\bibinfo {year} {2009})},\ \bibinfo {note}
  {(chapter III. A)}\BibitemShut {NoStop}%
\bibitem [{\citenamefont {Spalek}\ \emph {et~al.}(1987)\citenamefont {Spalek},
  \citenamefont {Datta},\ and\ \citenamefont {Honig}}]{Spalek1987}%
  \BibitemOpen
  \bibfield  {author} {\bibinfo {author} {\bibfnamefont {J.}~\bibnamefont
  {Spalek}}, \bibinfo {author} {\bibfnamefont {A.}~\bibnamefont {Datta}}, \
  and\ \bibinfo {author} {\bibfnamefont {J.~M.}\ \bibnamefont {Honig}},\ }\href
  {\doibase 10.1103/PhysRevLett.59.728} {\bibfield  {journal} {\bibinfo
  {journal} {Phys. Rev. Lett.}\ }\textbf {\bibinfo {volume} {59}},\ \bibinfo
  {pages} {728} (\bibinfo {year} {1987})}\BibitemShut {NoStop}%
\bibitem [{\citenamefont {Spa\l{}ek}(2006)}]{SpalekPhysStatSol2006}%
  \BibitemOpen
  \bibfield  {author} {\bibinfo {author} {\bibfnamefont {J.}~\bibnamefont
  {Spa\l{}ek}},\ }\href {\doibase 10.1002/pssb.200562526} {\bibfield  {journal}
  {\bibinfo  {journal} {physica status solidi (b)}\ }\textbf {\bibinfo {volume}
  {243}},\ \bibinfo {pages} {78} (\bibinfo {year} {2006})}\BibitemShut
  {NoStop}%
\bibitem [{\citenamefont {Saxena}\ \emph {et~al.}(2000)\citenamefont {Saxena},
  \citenamefont {Agarwal}, \citenamefont {Ahilan}, \citenamefont {Grosche},
  \citenamefont {Haselwimmer}, \citenamefont {Steiner}, \citenamefont {Pugh},
  \citenamefont {Walker}, \citenamefont {Julian}, \citenamefont {Monthoux},
  \citenamefont {Lonzarich}, \citenamefont {Huxley}, \citenamefont {Sheikin},
  \citenamefont {Braithwaite},\ and\ \citenamefont {Flouquet}}]{Saxena2000}%
  \BibitemOpen
  \bibfield  {author} {\bibinfo {author} {\bibfnamefont {S.~S.}\ \bibnamefont
  {Saxena}}, \bibinfo {author} {\bibfnamefont {P.}~\bibnamefont {Agarwal}},
  \bibinfo {author} {\bibfnamefont {K.}~\bibnamefont {Ahilan}}, \bibinfo
  {author} {\bibfnamefont {F.~M.}\ \bibnamefont {Grosche}}, \bibinfo {author}
  {\bibfnamefont {R.~K.~W.}\ \bibnamefont {Haselwimmer}}, \bibinfo {author}
  {\bibfnamefont {M.~J.}\ \bibnamefont {Steiner}}, \bibinfo {author}
  {\bibfnamefont {E.}~\bibnamefont {Pugh}}, \bibinfo {author} {\bibfnamefont
  {I.~R.}\ \bibnamefont {Walker}}, \bibinfo {author} {\bibfnamefont {S.~R.}\
  \bibnamefont {Julian}}, \bibinfo {author} {\bibfnamefont {P.}~\bibnamefont
  {Monthoux}}, \bibinfo {author} {\bibfnamefont {G.~G.}\ \bibnamefont
  {Lonzarich}}, \bibinfo {author} {\bibfnamefont {A.}~\bibnamefont {Huxley}},
  \bibinfo {author} {\bibfnamefont {I.}~\bibnamefont {Sheikin}}, \bibinfo
  {author} {\bibfnamefont {D.}~\bibnamefont {Braithwaite}}, \ and\ \bibinfo
  {author} {\bibfnamefont {J.}~\bibnamefont {Flouquet}},\ }\href {\doibase
  10.1038/35020500} {\bibfield  {journal} {\bibinfo  {journal} {Nature}\
  }\textbf {\bibinfo {volume} {406}},\ \bibinfo {pages} {587} (\bibinfo {year}
  {2000})}\BibitemShut {NoStop}%
\bibitem [{\citenamefont {Pfleiderer}\ and\ \citenamefont
  {Huxley}(2002)}]{Pfleiderer2002}%
  \BibitemOpen
  \bibfield  {author} {\bibinfo {author} {\bibfnamefont {C.}~\bibnamefont
  {Pfleiderer}}\ and\ \bibinfo {author} {\bibfnamefont {A.~D.}\ \bibnamefont
  {Huxley}},\ }\href {\doibase 10.1103/PhysRevLett.89.147005} {\bibfield
  {journal} {\bibinfo  {journal} {Phys. Rev. Lett.}\ }\textbf {\bibinfo
  {volume} {89}},\ \bibinfo {pages} {147005} (\bibinfo {year}
  {2002})}\BibitemShut {NoStop}%
\bibitem [{\citenamefont {Taufour}\ \emph {et~al.}(2010)\citenamefont
  {Taufour}, \citenamefont {Aoki}, \citenamefont {Knebel},\ and\ \citenamefont
  {Flouquet}}]{Taufour2010}%
  \BibitemOpen
  \bibfield  {author} {\bibinfo {author} {\bibfnamefont {V.}~\bibnamefont
  {Taufour}}, \bibinfo {author} {\bibfnamefont {D.}~\bibnamefont {Aoki}},
  \bibinfo {author} {\bibfnamefont {G.}~\bibnamefont {Knebel}}, \ and\ \bibinfo
  {author} {\bibfnamefont {J.}~\bibnamefont {Flouquet}},\ }\href {\doibase
  10.1103/PhysRevLett.105.217201} {\bibfield  {journal} {\bibinfo  {journal}
  {Phys. Rev. Lett.}\ }\textbf {\bibinfo {volume} {105}},\ \bibinfo {pages}
  {217201} (\bibinfo {year} {2010})}\BibitemShut {NoStop}%
\bibitem [{\citenamefont {Kotegawa}\ \emph {et~al.}(2011)\citenamefont
  {Kotegawa}, \citenamefont {Taufour}, \citenamefont {Aoki}, \citenamefont
  {Knebel},\ and\ \citenamefont {Flouquet}}]{Kotegawa2011}%
  \BibitemOpen
  \bibfield  {author} {\bibinfo {author} {\bibfnamefont {H.}~\bibnamefont
  {Kotegawa}}, \bibinfo {author} {\bibfnamefont {V.}~\bibnamefont {Taufour}},
  \bibinfo {author} {\bibfnamefont {D.}~\bibnamefont {Aoki}}, \bibinfo {author}
  {\bibfnamefont {G.}~\bibnamefont {Knebel}}, \ and\ \bibinfo {author}
  {\bibfnamefont {J.}~\bibnamefont {Flouquet}},\ }\href {\doibase
  10.1143/JPSJ.80.083703} {\bibfield  {journal} {\bibinfo  {journal} {J. Phys.
  Soc. Jpn.}\ }\textbf {\bibinfo {volume} {80}},\ \bibinfo {pages} {083703}
  (\bibinfo {year} {2011})}\BibitemShut {NoStop}%
\bibitem [{\citenamefont {Huxley}\ \emph {et~al.}(2001)\citenamefont {Huxley},
  \citenamefont {Sheikin}, \citenamefont {Ressouche}, \citenamefont
  {Kernavanois}, \citenamefont {Braithwaite}, \citenamefont {Calemczuk},\ and\
  \citenamefont {Flouquet}}]{Huxley2001}%
  \BibitemOpen
  \bibfield  {author} {\bibinfo {author} {\bibfnamefont {A.}~\bibnamefont
  {Huxley}}, \bibinfo {author} {\bibfnamefont {I.}~\bibnamefont {Sheikin}},
  \bibinfo {author} {\bibfnamefont {E.}~\bibnamefont {Ressouche}}, \bibinfo
  {author} {\bibfnamefont {N.}~\bibnamefont {Kernavanois}}, \bibinfo {author}
  {\bibfnamefont {D.}~\bibnamefont {Braithwaite}}, \bibinfo {author}
  {\bibfnamefont {R.}~\bibnamefont {Calemczuk}}, \ and\ \bibinfo {author}
  {\bibfnamefont {J.}~\bibnamefont {Flouquet}},\ }\href {\doibase
  10.1103/PhysRevB.63.144519} {\bibfield  {journal} {\bibinfo  {journal} {Phys.
  Rev. B}\ }\textbf {\bibinfo {volume} {63}},\ \bibinfo {pages} {144519}
  (\bibinfo {year} {2001})}\BibitemShut {NoStop}%
\bibitem [{\citenamefont {Kirkpatrick}\ \emph {et~al.}(2001)\citenamefont
  {Kirkpatrick}, \citenamefont {Belitz}, \citenamefont {Vojta},\ and\
  \citenamefont {Narayanan}}]{Kirkpatrick2001}%
  \BibitemOpen
  \bibfield  {author} {\bibinfo {author} {\bibfnamefont {T.~R.}\ \bibnamefont
  {Kirkpatrick}}, \bibinfo {author} {\bibfnamefont {D.}~\bibnamefont {Belitz}},
  \bibinfo {author} {\bibfnamefont {T.}~\bibnamefont {Vojta}}, \ and\ \bibinfo
  {author} {\bibfnamefont {R.}~\bibnamefont {Narayanan}},\ }\href {\doibase
  10.1103/PhysRevLett.87.127003} {\bibfield  {journal} {\bibinfo  {journal}
  {Phys. Rev. Lett.}\ }\textbf {\bibinfo {volume} {87}},\ \bibinfo {pages}
  {127003} (\bibinfo {year} {2001})}\BibitemShut {NoStop}%
\bibitem [{\citenamefont {Machida}\ and\ \citenamefont
  {Ohmi}(2001)}]{Machida2001}%
  \BibitemOpen
  \bibfield  {author} {\bibinfo {author} {\bibfnamefont {K.}~\bibnamefont
  {Machida}}\ and\ \bibinfo {author} {\bibfnamefont {T.}~\bibnamefont {Ohmi}},\
  }\href {\doibase 10.1103/PhysRevLett.86.850} {\bibfield  {journal} {\bibinfo
  {journal} {Phys. Rev. Lett.}\ }\textbf {\bibinfo {volume} {86}},\ \bibinfo
  {pages} {850} (\bibinfo {year} {2001})}\BibitemShut {NoStop}%
\bibitem [{\citenamefont {Abrikosov}(2001)}]{Abrikosov2001}%
  \BibitemOpen
  \bibfield  {author} {\bibinfo {author} {\bibfnamefont {A.~A.}\ \bibnamefont
  {Abrikosov}},\ }\href@noop {} {\bibfield  {journal} {\bibinfo  {journal} {J.
  Phys: Condens. Matter}\ }\textbf {\bibinfo {volume} {13}},\ \bibinfo {pages}
  {L943} (\bibinfo {year} {2001})}\BibitemShut {NoStop}%
\bibitem [{\citenamefont {Sa}(2002)}]{Sa2002}%
  \BibitemOpen
  \bibfield  {author} {\bibinfo {author} {\bibfnamefont {D.}~\bibnamefont
  {Sa}},\ }\href {\doibase 10.1103/PhysRevB.66.140505} {\bibfield  {journal}
  {\bibinfo  {journal} {Phys. Rev. B}\ }\textbf {\bibinfo {volume} {66}},\
  \bibinfo {pages} {140505} (\bibinfo {year} {2002})}\BibitemShut {NoStop}%
\bibitem [{\citenamefont {Sandeman}\ \emph {et~al.}(2003)\citenamefont
  {Sandeman}, \citenamefont {Lonzarich},\ and\ \citenamefont
  {Schofield}}]{Sandeman2003}%
  \BibitemOpen
  \bibfield  {author} {\bibinfo {author} {\bibfnamefont {K.~G.}\ \bibnamefont
  {Sandeman}}, \bibinfo {author} {\bibfnamefont {G.~G.}\ \bibnamefont
  {Lonzarich}}, \ and\ \bibinfo {author} {\bibfnamefont {A.~J.}\ \bibnamefont
  {Schofield}},\ }\href {\doibase 10.1103/PhysRevLett.90.167005} {\bibfield
  {journal} {\bibinfo  {journal} {Phys. Rev. Lett.}\ }\textbf {\bibinfo
  {volume} {90}},\ \bibinfo {pages} {167005} (\bibinfo {year}
  {2003})}\BibitemShut {NoStop}%
\bibitem [{\citenamefont {Belitz}\ \emph {et~al.}(2005)\citenamefont {Belitz},
  \citenamefont {Kirkpatrick},\ and\ \citenamefont
  {Rollb\"uhler}}]{Kirkpatrick2005}%
  \BibitemOpen
  \bibfield  {author} {\bibinfo {author} {\bibfnamefont {D.}~\bibnamefont
  {Belitz}}, \bibinfo {author} {\bibfnamefont {T.~R.}\ \bibnamefont
  {Kirkpatrick}}, \ and\ \bibinfo {author} {\bibfnamefont {J.}~\bibnamefont
  {Rollb\"uhler}},\ }\href {\doibase 10.1103/PhysRevLett.94.247205} {\bibfield
  {journal} {\bibinfo  {journal} {Phys. Rev. Lett.}\ }\textbf {\bibinfo
  {volume} {94}},\ \bibinfo {pages} {247205} (\bibinfo {year}
  {2005})}\BibitemShut {NoStop}%
\bibitem [{\citenamefont {Hardy}\ \emph {et~al.}(2009)\citenamefont {Hardy},
  \citenamefont {Meingast}, \citenamefont {Taufour}, \citenamefont {Flouquet},
  \citenamefont {v.~L\"ohneysen}, \citenamefont {Fisher}, \citenamefont
  {Phillips}, \citenamefont {Huxley},\ and\ \citenamefont
  {Lashley}}]{Hardy2009}%
  \BibitemOpen
  \bibfield  {author} {\bibinfo {author} {\bibfnamefont {F.}~\bibnamefont
  {Hardy}}, \bibinfo {author} {\bibfnamefont {C.}~\bibnamefont {Meingast}},
  \bibinfo {author} {\bibfnamefont {V.}~\bibnamefont {Taufour}}, \bibinfo
  {author} {\bibfnamefont {J.}~\bibnamefont {Flouquet}}, \bibinfo {author}
  {\bibfnamefont {H.}~\bibnamefont {v.~L\"ohneysen}}, \bibinfo {author}
  {\bibfnamefont {R.~A.}\ \bibnamefont {Fisher}}, \bibinfo {author}
  {\bibfnamefont {N.~E.}\ \bibnamefont {Phillips}}, \bibinfo {author}
  {\bibfnamefont {A.}~\bibnamefont {Huxley}}, \ and\ \bibinfo {author}
  {\bibfnamefont {J.~C.}\ \bibnamefont {Lashley}},\ }\href {\doibase
  10.1103/PhysRevB.80.174521} {\bibfield  {journal} {\bibinfo  {journal} {Phys.
  Rev. B}\ }\textbf {\bibinfo {volume} {80}},\ \bibinfo {pages} {174521}
  (\bibinfo {year} {2009})}\BibitemShut {NoStop}%
\bibitem [{\citenamefont {Wysoki\ifmmode~\acute{n}\else \'{n}\fi{}ski}\ \emph
  {et~al.}(2014)\citenamefont {Wysoki\ifmmode~\acute{n}\else \'{n}\fi{}ski},
  \citenamefont {Abram},\ and\ \citenamefont {Spa\l{}ek}}]{Rapid}%
  \BibitemOpen
  \bibfield  {author} {\bibinfo {author} {\bibfnamefont {M.~M.}\ \bibnamefont
  {Wysoki\ifmmode~\acute{n}\else \'{n}\fi{}ski}}, \bibinfo {author}
  {\bibfnamefont {M.}~\bibnamefont {Abram}}, \ and\ \bibinfo {author}
  {\bibfnamefont {J.}~\bibnamefont {Spa\l{}ek}},\ }\href {\doibase
  10.1103/PhysRevB.90.081114} {\bibfield  {journal} {\bibinfo  {journal} {Phys.
  Rev. B}\ }\textbf {\bibinfo {volume} {90}},\ \bibinfo {pages} {081114(R)}
  (\bibinfo {year} {2014})}\BibitemShut {NoStop}%
\bibitem [{\citenamefont {J\k{e}drak}\ \emph {et~al.}()\citenamefont
  {J\k{e}drak}, \citenamefont {Kaczmarczyk},\ and\ \citenamefont
  {Spa{\l}ek}}]{sga}%
  \BibitemOpen
  \bibfield  {author} {\bibinfo {author} {\bibfnamefont {J.}~\bibnamefont
  {J\k{e}drak}}, \bibinfo {author} {\bibfnamefont {J.}~\bibnamefont
  {Kaczmarczyk}}, \ and\ \bibinfo {author} {\bibfnamefont {J.}~\bibnamefont
  {Spa{\l}ek}},\ }\href@noop {} {\bibinfo  {journal} {arXiv:1008.0021}\
  }\BibitemShut {NoStop}%
\bibitem [{\citenamefont {J\k{e}drak}\ and\ \citenamefont
  {Spa{\l}ek}(2011)}]{Jedrak2011}%
  \BibitemOpen
\bibfield  {journal} {  }\bibfield  {author} {\bibinfo {author} {\bibfnamefont
  {J.}~\bibnamefont {J\k{e}drak}}\ and\ \bibinfo {author} {\bibfnamefont
  {J.}~\bibnamefont {Spa{\l}ek}},\ }\href {\doibase 10.1103/PhysRevB.83.104512}
  {\bibfield  {journal} {\bibinfo  {journal} {Phys. Rev. B}\ }\textbf {\bibinfo
  {volume} {83}},\ \bibinfo {pages} {104512} (\bibinfo {year}
  {2011})}\BibitemShut {NoStop}%
\bibitem [{\citenamefont {Kaczmarczyk}\ and\ \citenamefont
  {Spa\l{}ek}(2011)}]{Kaczmarczyk2011}%
  \BibitemOpen
  \bibfield  {author} {\bibinfo {author} {\bibfnamefont {J.}~\bibnamefont
  {Kaczmarczyk}}\ and\ \bibinfo {author} {\bibfnamefont {J.}~\bibnamefont
  {Spa\l{}ek}},\ }\href {\doibase 10.1103/PhysRevB.84.125140} {\bibfield
  {journal} {\bibinfo  {journal} {Phys. Rev. B}\ }\textbf {\bibinfo {volume}
  {84}},\ \bibinfo {pages} {125140} (\bibinfo {year} {2011})}\BibitemShut
  {NoStop}%
\bibitem [{\citenamefont {Howczak}\ \emph {et~al.}(2013)\citenamefont
  {Howczak}, \citenamefont {Kaczmarczyk},\ and\ \citenamefont
  {Spa{\l}ek}}]{Howczak2013}%
  \BibitemOpen
  \bibfield  {author} {\bibinfo {author} {\bibfnamefont {O.}~\bibnamefont
  {Howczak}}, \bibinfo {author} {\bibfnamefont {J.}~\bibnamefont
  {Kaczmarczyk}}, \ and\ \bibinfo {author} {\bibfnamefont {J.}~\bibnamefont
  {Spa{\l}ek}},\ }\href {\doibase 10.1002/pssb.201200774} {\bibfield  {journal}
  {\bibinfo  {journal} {Phys. Stat. Solidi (b)}\ }\textbf {\bibinfo {volume}
  {250}},\ \bibinfo {pages} {609} (\bibinfo {year} {2013})}\BibitemShut
  {NoStop}%
\bibitem [{\citenamefont {K\k{a}dzielawa}\ \emph {et~al.}(2013)\citenamefont
  {K\k{a}dzielawa}, \citenamefont {Spa{\l}ek}, \citenamefont {Kurzyk},\ and\
  \citenamefont {W\'{o}jcik}}]{Kadzielawa2013}%
  \BibitemOpen
  \bibfield  {author} {\bibinfo {author} {\bibfnamefont {A.~P.}\ \bibnamefont
  {K\k{a}dzielawa}}, \bibinfo {author} {\bibfnamefont {J.}~\bibnamefont
  {Spa{\l}ek}}, \bibinfo {author} {\bibfnamefont {J.}~\bibnamefont {Kurzyk}}, \
  and\ \bibinfo {author} {\bibfnamefont {W.}~\bibnamefont {W\'{o}jcik}},\
  }\href@noop {} {\bibfield  {journal} {\bibinfo  {journal} {Eur. Phys. J. B}\
  }\textbf {\bibinfo {volume} {86}},\ \bibinfo {pages} {252} (\bibinfo {year}
  {2013})}\BibitemShut {NoStop}%
\bibitem [{\citenamefont {Abram}\ \emph {et~al.}(2013)\citenamefont {Abram},
  \citenamefont {Kaczmarczyk}, \citenamefont {J\k{e}drak},\ and\ \citenamefont
  {Spa\l{}ek}}]{Abram2013}%
  \BibitemOpen
  \bibfield  {author} {\bibinfo {author} {\bibfnamefont {M.}~\bibnamefont
  {Abram}}, \bibinfo {author} {\bibfnamefont {J.}~\bibnamefont {Kaczmarczyk}},
  \bibinfo {author} {\bibfnamefont {J.}~\bibnamefont {J\k{e}drak}}, \ and\
  \bibinfo {author} {\bibfnamefont {J.}~\bibnamefont {Spa\l{}ek}},\ }\href
  {\doibase 10.1103/PhysRevB.88.094502} {\bibfield  {journal} {\bibinfo
  {journal} {Phys. Rev. B}\ }\textbf {\bibinfo {volume} {88}},\ \bibinfo
  {pages} {094502} (\bibinfo {year} {2013})}\BibitemShut {NoStop}%
\bibitem [{\citenamefont {Zegrodnik}\ \emph {et~al.}(2013)\citenamefont
  {Zegrodnik}, \citenamefont {Spa\l{}ek},\ and\ \citenamefont
  {B\"unemann}}]{Zegrodnik2013}%
  \BibitemOpen
  \bibfield  {author} {\bibinfo {author} {\bibfnamefont {M.}~\bibnamefont
  {Zegrodnik}}, \bibinfo {author} {\bibfnamefont {J.}~\bibnamefont
  {Spa\l{}ek}}, \ and\ \bibinfo {author} {\bibfnamefont {J.}~\bibnamefont
  {B\"unemann}},\ }\href {\doibase 10.1088/1367-2630/15/7/073050} {\bibfield
  {journal} {\bibinfo  {journal} {New J. Phys.}\ }\textbf {\bibinfo {volume}
  {15}},\ \bibinfo {pages} {073050} (\bibinfo {year} {2013})}\BibitemShut
  {NoStop}%
\bibitem [{\citenamefont {Wysoki\'nski}\ and\ \citenamefont
  {Spa{\l}ek}(2014)}]{Wysokinski2014}%
  \BibitemOpen
  \bibfield  {author} {\bibinfo {author} {\bibfnamefont {M.~M.}\ \bibnamefont
  {Wysoki\'nski}}\ and\ \bibinfo {author} {\bibfnamefont {J.}~\bibnamefont
  {Spa{\l}ek}},\ }\href {\doibase 10.1088/0953-8984/26/5/055601} {\bibfield
  {journal} {\bibinfo  {journal} {J. Phys.: Condens. Matter}\ }\textbf
  {\bibinfo {volume} {26}},\ \bibinfo {pages} {055601} (\bibinfo {year}
  {2014})}\BibitemShut {NoStop}%
\bibitem [{\citenamefont {Shick}\ and\ \citenamefont
  {Pickett}(2001)}]{Shick2001}%
  \BibitemOpen
  \bibfield  {author} {\bibinfo {author} {\bibfnamefont {A.~B.}\ \bibnamefont
  {Shick}}\ and\ \bibinfo {author} {\bibfnamefont {W.~E.}\ \bibnamefont
  {Pickett}},\ }\href {\doibase 10.1103/PhysRevLett.86.300} {\bibfield
  {journal} {\bibinfo  {journal} {Phys. Rev. Lett.}\ }\textbf {\bibinfo
  {volume} {86}},\ \bibinfo {pages} {300} (\bibinfo {year} {2001})}\BibitemShut
  {NoStop}%
\bibitem [{\citenamefont {Samsel-Czeka\l{}a}\ \emph {et~al.}(2011)\citenamefont
  {Samsel-Czeka\l{}a}, \citenamefont {Werwi\'nski}, \citenamefont {Szajek},
  \citenamefont {Che\l{}kowska},\ and\ \citenamefont {Tro\'c}}]{Samsel2011}%
  \BibitemOpen
  \bibfield  {author} {\bibinfo {author} {\bibfnamefont {M.}~\bibnamefont
  {Samsel-Czeka\l{}a}}, \bibinfo {author} {\bibfnamefont {M.}~\bibnamefont
  {Werwi\'nski}}, \bibinfo {author} {\bibfnamefont {A.}~\bibnamefont {Szajek}},
  \bibinfo {author} {\bibfnamefont {G.}~\bibnamefont {Che\l{}kowska}}, \ and\
  \bibinfo {author} {\bibfnamefont {R.}~\bibnamefont {Tro\'c}},\ }\href@noop {}
  {\bibfield  {journal} {\bibinfo  {journal} {Intermetallics}\ }\textbf
  {\bibinfo {volume} {19}},\ \bibinfo {pages} {1411} (\bibinfo {year}
  {2011})}\BibitemShut {NoStop}%
\bibitem [{\citenamefont {Tran}\ \emph {et~al.}(2004)\citenamefont {Tran},
  \citenamefont {Paschen}, \citenamefont {Tro\ifmmode~\acute{c}\else
  \'{c}\fi{}}, \citenamefont {Baenitz},\ and\ \citenamefont
  {Steglich}}]{Tran2004}%
  \BibitemOpen
  \bibfield  {author} {\bibinfo {author} {\bibfnamefont {V.~H.}\ \bibnamefont
  {Tran}}, \bibinfo {author} {\bibfnamefont {S.}~\bibnamefont {Paschen}},
  \bibinfo {author} {\bibfnamefont {R.}~\bibnamefont
  {Tro\ifmmode~\acute{c}\else \'{c}\fi{}}}, \bibinfo {author} {\bibfnamefont
  {M.}~\bibnamefont {Baenitz}}, \ and\ \bibinfo {author} {\bibfnamefont
  {F.}~\bibnamefont {Steglich}},\ }\href {\doibase 10.1103/PhysRevB.69.195314}
  {\bibfield  {journal} {\bibinfo  {journal} {Phys. Rev. B}\ }\textbf {\bibinfo
  {volume} {69}},\ \bibinfo {pages} {195314} (\bibinfo {year}
  {2004})}\BibitemShut {NoStop}%
\bibitem [{\citenamefont {Kernavanois}\ \emph {et~al.}(2001)\citenamefont
  {Kernavanois}, \citenamefont {Grenier}, \citenamefont {Huxley}, \citenamefont
  {Ressouche}, \citenamefont {Sanchez},\ and\ \citenamefont
  {Flouquet}}]{Kernavanois2001}%
  \BibitemOpen
  \bibfield  {author} {\bibinfo {author} {\bibfnamefont {N.}~\bibnamefont
  {Kernavanois}}, \bibinfo {author} {\bibfnamefont {B.}~\bibnamefont
  {Grenier}}, \bibinfo {author} {\bibfnamefont {A.}~\bibnamefont {Huxley}},
  \bibinfo {author} {\bibfnamefont {E.}~\bibnamefont {Ressouche}}, \bibinfo
  {author} {\bibfnamefont {J.~P.}\ \bibnamefont {Sanchez}}, \ and\ \bibinfo
  {author} {\bibfnamefont {J.}~\bibnamefont {Flouquet}},\ }\href {\doibase
  10.1103/PhysRevB.64.174509} {\bibfield  {journal} {\bibinfo  {journal} {Phys.
  Rev. B}\ }\textbf {\bibinfo {volume} {64}},\ \bibinfo {pages} {174509}
  (\bibinfo {year} {2001})}\BibitemShut {NoStop}%
\bibitem [{\citenamefont {Doradzi\ifmmode~\acute{n}\else \'{n}\fi{}ski}\ and\
  \citenamefont {Spa\l{}ek}(1997)}]{Doradzinski1997}%
  \BibitemOpen
  \bibfield  {author} {\bibinfo {author} {\bibfnamefont {R.}~\bibnamefont
  {Doradzi\ifmmode~\acute{n}\else \'{n}\fi{}ski}}\ and\ \bibinfo {author}
  {\bibfnamefont {J.}~\bibnamefont {Spa\l{}ek}},\ }\href {\doibase
  10.1103/PhysRevB.56.R14239} {\bibfield  {journal} {\bibinfo  {journal} {Phys.
  Rev. B}\ }\textbf {\bibinfo {volume} {56}},\ \bibinfo {pages} {R14239}
  (\bibinfo {year} {1997})}\BibitemShut {NoStop}%
\bibitem [{\citenamefont {Doradzi\ifmmode~\acute{n}\else \'{n}\fi{}ski}\ and\
  \citenamefont {Spa\l{}ek}(1998)}]{Doradzinski1998}%
  \BibitemOpen
  \bibfield  {author} {\bibinfo {author} {\bibfnamefont {R.}~\bibnamefont
  {Doradzi\ifmmode~\acute{n}\else \'{n}\fi{}ski}}\ and\ \bibinfo {author}
  {\bibfnamefont {J.}~\bibnamefont {Spa\l{}ek}},\ }\href {\doibase
  10.1103/PhysRevB.58.3293} {\bibfield  {journal} {\bibinfo  {journal} {Phys.
  Rev. B}\ }\textbf {\bibinfo {volume} {58}},\ \bibinfo {pages} {3293}
  (\bibinfo {year} {1998})}\BibitemShut {NoStop}%
\bibitem [{\citenamefont {Howczak}\ and\ \citenamefont
  {Spa\l{}ek}(2012)}]{Howczak2012}%
  \BibitemOpen
  \bibfield  {author} {\bibinfo {author} {\bibfnamefont {O.}~\bibnamefont
  {Howczak}}\ and\ \bibinfo {author} {\bibfnamefont {J.}~\bibnamefont
  {Spa\l{}ek}},\ }\href {\doibase 10.1088/0953-8984/24/20/205602} {\bibfield
  {journal} {\bibinfo  {journal} {J. Phys: Condens. Matter}\ }\textbf {\bibinfo
  {volume} {24}},\ \bibinfo {pages} {205602} (\bibinfo {year}
  {2012})}\BibitemShut {NoStop}%
\bibitem [{\citenamefont {Kubo}(2013)}]{Kubo2013}%
  \BibitemOpen
  \bibfield  {author} {\bibinfo {author} {\bibfnamefont {K.}~\bibnamefont
  {Kubo}},\ }\href {\doibase 10.1103/PhysRevB.87.195127} {\bibfield  {journal}
  {\bibinfo  {journal} {Phys. Rev. B}\ }\textbf {\bibinfo {volume} {87}},\
  \bibinfo {pages} {195127} (\bibinfo {year} {2013})}\BibitemShut {NoStop}%
\bibitem [{\citenamefont {Kotliar}\ and\ \citenamefont
  {Ruckenstein}(1986)}]{Kotliar1986}%
  \BibitemOpen
  \bibfield  {author} {\bibinfo {author} {\bibfnamefont {G.}~\bibnamefont
  {Kotliar}}\ and\ \bibinfo {author} {\bibfnamefont {A.~E.}\ \bibnamefont
  {Ruckenstein}},\ }\href {\doibase 10.1103/PhysRevLett.57.1362} {\bibfield
  {journal} {\bibinfo  {journal} {Phys. Rev. Lett.}\ }\textbf {\bibinfo
  {volume} {57}},\ \bibinfo {pages} {1362} (\bibinfo {year}
  {1986})}\BibitemShut {NoStop}%
\bibitem [{\citenamefont {Dorin}\ and\ \citenamefont
  {Schlottmann}(1992)}]{Dorin1992}%
  \BibitemOpen
  \bibfield  {author} {\bibinfo {author} {\bibfnamefont {V.}~\bibnamefont
  {Dorin}}\ and\ \bibinfo {author} {\bibfnamefont {P.}~\bibnamefont
  {Schlottmann}},\ }\href {\doibase 10.1103/PhysRevB.46.10800} {\bibfield
  {journal} {\bibinfo  {journal} {Phys. Rev. B}\ }\textbf {\bibinfo {volume}
  {46}},\ \bibinfo {pages} {10800} (\bibinfo {year} {1992})}\BibitemShut
  {NoStop}%
\bibitem [{\citenamefont {Terashima}\ \emph {et~al.}(2001)\citenamefont
  {Terashima}, \citenamefont {Matsumoto}, \citenamefont {Terakura},
  \citenamefont {Uji}, \citenamefont {Kimura}, \citenamefont {Endo},
  \citenamefont {Komatsubara},\ and\ \citenamefont {Aoki}}]{Terashima2001}%
  \BibitemOpen
  \bibfield  {author} {\bibinfo {author} {\bibfnamefont {T.}~\bibnamefont
  {Terashima}}, \bibinfo {author} {\bibfnamefont {T.}~\bibnamefont
  {Matsumoto}}, \bibinfo {author} {\bibfnamefont {C.}~\bibnamefont {Terakura}},
  \bibinfo {author} {\bibfnamefont {S.}~\bibnamefont {Uji}}, \bibinfo {author}
  {\bibfnamefont {N.}~\bibnamefont {Kimura}}, \bibinfo {author} {\bibfnamefont
  {M.}~\bibnamefont {Endo}}, \bibinfo {author} {\bibfnamefont {T.}~\bibnamefont
  {Komatsubara}}, \ and\ \bibinfo {author} {\bibfnamefont {H.}~\bibnamefont
  {Aoki}},\ }\href {\doibase 10.1103/PhysRevLett.87.166401} {\bibfield
  {journal} {\bibinfo  {journal} {Phys. Rev. Lett.}\ }\textbf {\bibinfo
  {volume} {87}},\ \bibinfo {pages} {166401} (\bibinfo {year}
  {2001})}\BibitemShut {NoStop}%
\bibitem [{\citenamefont {Settai}\ \emph {et~al.}(2002)\citenamefont {Settai},
  \citenamefont {Nakashima}, \citenamefont {Araki}, \citenamefont {Haga},
  \citenamefont {Kobayashi}, \citenamefont {Tateiwa}, \citenamefont
  {Yamagami},\ and\ \citenamefont {Onuki}}]{Settai2002}%
  \BibitemOpen
  \bibfield  {author} {\bibinfo {author} {\bibfnamefont {R.}~\bibnamefont
  {Settai}}, \bibinfo {author} {\bibfnamefont {M.}~\bibnamefont {Nakashima}},
  \bibinfo {author} {\bibfnamefont {S.}~\bibnamefont {Araki}}, \bibinfo
  {author} {\bibfnamefont {Y.}~\bibnamefont {Haga}}, \bibinfo {author}
  {\bibfnamefont {T.~C.}\ \bibnamefont {Kobayashi}}, \bibinfo {author}
  {\bibfnamefont {N.}~\bibnamefont {Tateiwa}}, \bibinfo {author} {\bibfnamefont
  {H.}~\bibnamefont {Yamagami}}, \ and\ \bibinfo {author} {\bibfnamefont
  {Y.}~\bibnamefont {Onuki}},\ }\href {\doibase 10.1088/0953-8984/14/1/104}
  {\bibfield  {journal} {\bibinfo  {journal} {J. Phys: Condens. Matter}\
  }\textbf {\bibinfo {volume} {14}},\ \bibinfo {pages} {L29} (\bibinfo {year}
  {2002})}\BibitemShut {NoStop}%
\bibitem [{\citenamefont {Rice}\ and\ \citenamefont {Ueda}(1985)}]{Rice1985}%
  \BibitemOpen
  \bibfield  {author} {\bibinfo {author} {\bibfnamefont {T.~M.}\ \bibnamefont
  {Rice}}\ and\ \bibinfo {author} {\bibfnamefont {K.}~\bibnamefont {Ueda}},\
  }\href {\doibase 10.1103/PhysRevLett.55.995} {\bibfield  {journal} {\bibinfo
  {journal} {Phys. Rev. Lett.}\ }\textbf {\bibinfo {volume} {55}},\ \bibinfo
  {pages} {995} (\bibinfo {year} {1985})}\BibitemShut {NoStop}%
\bibitem [{\citenamefont {Fazekas}\ and\ \citenamefont
  {Brandow}(1987)}]{Fazekas1987}%
  \BibitemOpen
  \bibfield  {author} {\bibinfo {author} {\bibfnamefont {P.}~\bibnamefont
  {Fazekas}}\ and\ \bibinfo {author} {\bibfnamefont {B.~H.}\ \bibnamefont
  {Brandow}},\ }\href@noop {} {\bibfield  {journal} {\bibinfo  {journal} {Phys.
  Scr.}\ }\textbf {\bibinfo {volume} {36}},\ \bibinfo {pages} {809} (\bibinfo
  {year} {1987})}\BibitemShut {NoStop}%
\bibitem [{\citenamefont {Wang}\ \emph {et~al.}(2010)\citenamefont {Wang},
  \citenamefont {He}, \citenamefont {Wang}, \citenamefont {Wang}, \citenamefont
  {Wang},\ and\ \citenamefont {Zhang}}]{Wang2010}%
  \BibitemOpen
  \bibfield  {author} {\bibinfo {author} {\bibfnamefont {W.-S.}\ \bibnamefont
  {Wang}}, \bibinfo {author} {\bibfnamefont {X.-M.}\ \bibnamefont {He}},
  \bibinfo {author} {\bibfnamefont {D.}~\bibnamefont {Wang}}, \bibinfo {author}
  {\bibfnamefont {Q.-H.}\ \bibnamefont {Wang}}, \bibinfo {author}
  {\bibfnamefont {Z.~D.}\ \bibnamefont {Wang}}, \ and\ \bibinfo {author}
  {\bibfnamefont {F.~C.}\ \bibnamefont {Zhang}},\ }\href {\doibase
  10.1103/PhysRevB.82.125105} {\bibfield  {journal} {\bibinfo  {journal} {Phys.
  Rev. B}\ }\textbf {\bibinfo {volume} {82}},\ \bibinfo {pages} {125105}
  (\bibinfo {year} {2010})}\BibitemShut {NoStop}%
\bibitem [{\citenamefont {Sandri}\ \emph {et~al.}(2013)\citenamefont {Sandri},
  \citenamefont {Capone},\ and\ \citenamefont {Fabrizio}}]{Fabrizio2013}%
  \BibitemOpen
  \bibfield  {author} {\bibinfo {author} {\bibfnamefont {M.}~\bibnamefont
  {Sandri}}, \bibinfo {author} {\bibfnamefont {M.}~\bibnamefont {Capone}}, \
  and\ \bibinfo {author} {\bibfnamefont {M.}~\bibnamefont {Fabrizio}},\ }\href
  {\doibase 10.1103/PhysRevB.87.205108} {\bibfield  {journal} {\bibinfo
  {journal} {Phys. Rev. B}\ }\textbf {\bibinfo {volume} {87}},\ \bibinfo
  {pages} {205108} (\bibinfo {year} {2013})}\BibitemShut {NoStop}%
\bibitem [{\citenamefont {Miyake}\ and\ \citenamefont
  {Watanabe}(2014)}]{Miyake2014}%
  \BibitemOpen
  \bibfield  {author} {\bibinfo {author} {\bibfnamefont {K.}~\bibnamefont
  {Miyake}}\ and\ \bibinfo {author} {\bibfnamefont {S.}~\bibnamefont
  {Watanabe}},\ }\href {\doibase 10.7566/JPSJ.83.061006} {\bibfield  {journal}
  {\bibinfo  {journal} {J. Phys. Soc. Jpn.}\ }\textbf {\bibinfo {volume}
  {83}},\ \bibinfo {pages} {061006} (\bibinfo {year} {2014})}\BibitemShut
  {NoStop}%
\bibitem [{\citenamefont {Roussev}\ and\ \citenamefont
  {Millis}(2001)}]{Millis2001}%
  \BibitemOpen
  \bibfield  {author} {\bibinfo {author} {\bibfnamefont {R.}~\bibnamefont
  {Roussev}}\ and\ \bibinfo {author} {\bibfnamefont {A.~J.}\ \bibnamefont
  {Millis}},\ }\href {\doibase 10.1103/PhysRevB.63.140504} {\bibfield
  {journal} {\bibinfo  {journal} {Phys. Rev. B}\ }\textbf {\bibinfo {volume}
  {63}},\ \bibinfo {pages} {140504} (\bibinfo {year} {2001})}\BibitemShut
  {NoStop}%
\bibitem [{\citenamefont {Fay}\ and\ \citenamefont {Appel}(1980)}]{Fay1980}%
  \BibitemOpen
  \bibfield  {author} {\bibinfo {author} {\bibfnamefont {D.}~\bibnamefont
  {Fay}}\ and\ \bibinfo {author} {\bibfnamefont {J.}~\bibnamefont {Appel}},\
  }\href {\doibase 10.1103/PhysRevB.22.3173} {\bibfield  {journal} {\bibinfo
  {journal} {Phys. Rev. B}\ }\textbf {\bibinfo {volume} {22}},\ \bibinfo
  {pages} {3173} (\bibinfo {year} {1980})}\BibitemShut {NoStop}%
\bibitem [{\citenamefont {Zegrodnik}\ \emph {et~al.}(2014)\citenamefont
  {Zegrodnik}, \citenamefont {B\"unemann},\ and\ \citenamefont
  {Spa\l{}ek}}]{Zegrodnik2014}%
  \BibitemOpen
  \bibfield  {author} {\bibinfo {author} {\bibfnamefont {M.}~\bibnamefont
  {Zegrodnik}}, \bibinfo {author} {\bibfnamefont {J.}~\bibnamefont
  {B\"unemann}}, \ and\ \bibinfo {author} {\bibfnamefont {J.}~\bibnamefont
  {Spa\l{}ek}},\ }\href {\doibase 10.1088/1367-2630/16/3/033001} {\bibfield
  {journal} {\bibinfo  {journal} {New J. Phys.}\ }\textbf {\bibinfo {volume}
  {16}},\ \bibinfo {pages} {033001} (\bibinfo {year} {2014})}\BibitemShut
  {NoStop}%
\bibitem [{\citenamefont {Spa\l{}ek}\ and\ \citenamefont
  {Zegrodnik}(2013)}]{Zegrodnik2013JPCM}%
  \BibitemOpen
  \bibfield  {author} {\bibinfo {author} {\bibfnamefont {J.}~\bibnamefont
  {Spa\l{}ek}}\ and\ \bibinfo {author} {\bibfnamefont {M.}~\bibnamefont
  {Zegrodnik}},\ }\href@noop {} {\bibfield  {journal} {\bibinfo  {journal} {J.
  Phys.: Condens. Matter}\ }\textbf {\bibinfo {volume} {25}},\ \bibinfo {pages}
  {435601} (\bibinfo {year} {2013})}\BibitemShut {NoStop}%
\end{thebibliography}
 %
\end{document}